\documentclass[3p,twocolumn]{elsarticle}
\usepackage{amsmath,amssymb,amsfonts}
\usepackage{subcaption}
\usepackage{algorithmic}
\usepackage{graphicx}
\usepackage{bm}
\usepackage{textcomp}
\usepackage{xcolor}
\usepackage{footnote}
\usepackage{hyperref}
\usepackage{multirow}
\usepackage{pifont}
\makesavenoteenv{tabular}
\journal{Journal of \LaTeX\ Templates}

\begin{document}
	
	\begin{frontmatter}
		
		\title{Accent Recognition with Hybrid {Phonetic Features}}

		
		\author[mymainaddress]{Zhan Zhang}
		\ead{zhan\_zhang@zju.edu.cn}
		\author[mymainaddress]{Xi Chen}
		\ead{chen\_\_xi@zju.edu.cn}
		\author[mymainaddress]{Yuehai Wang\corref{mycorrespondingauthor}}
		\cortext[mycorrespondingauthor]{Corresponding author}
		\ead{wyuehai@zju.edu.cn}
		
		\author[mymainaddress]{Jianyi Yang}
		\ead{yangjy@zju.edu.cn}
		
		\address[mymainaddress]{ Department of Information and Electronic Engineering, Zhejiang University, China}

		\begin{abstract}
			The performance of voice-controlled systems is usually influenced by accented speech. To make these systems more robust, the frontend accent recognition (AR) technologies have received increased attention in recent years.
			As accent is a high-level abstract feature that has a profound relationship with the language knowledge, AR is more challenging than other language-agnostic audio classification tasks. In this paper, we use an auxiliary automatic speech recognition (ASR) task to extract language-related {phonetic} features. Furthermore, we propose a hybrid structure that incorporates the embeddings of both a {fixed} acoustic model and a {trainable} acoustic model, making the language-related acoustic feature more robust. We conduct several experiments on the Accented English Speech Recognition Challenge (AESRC) 2020 dataset. The results demonstrate that our approach can obtain a 6.57\% relative improvement on the validation set. We also get a 7.28\% relative improvement on the final test set for this competition, showing the merits of the proposed method.
			

		\end{abstract}
		
		
	\end{frontmatter}
	

	\section{Introduction}
	\label{sec:intro}
	With the quick growth of voice-controlled systems, speech-related technologies are becoming part of our daily life. However, the variability of speech poses a serious challenge to these technologies. Among the many factors that can influence the speech variability, accent is a typical one that {will cause degradation in recognition accuracy\cite{Chu2021410} \cite{Huang2001}\cite{Levis2011}}.
	
	
	Accent is a diverse pronouncing behavior under certain languages, which can be influenced by social environment, education, residential zone, and so on. As analyzed in \cite{Viglino2019},  {English speakers }  are constructed by not only about 380 million natives, but also by close to 740 million non-native speakers. Influenced by their {native language}, the speakers may have a very wide variety of accents.
	
	To analyze the accent attribute in the collected speech and make the whole voice-controlled system more generalized, accent recognition (AR) or accent classification technologies can be applied to custom the downstream subsystems. Thus, AR technologies have received increased attention in recent years. 
	

	From the point of deep learning, as accent is an utterance-level attribute, AR is also a classification task that converts an audio sequence into a certain class.  In this respect, the audio classification tasks, including audio scene classification, speaker recognition, and AR, can share similar ideas on network structures. 	
	However, AR is a more challenging task. 
		
	{Generally, the acoustic scene classification or speaker recognition task can be finished by using certain low-level discriminative features. For example, speaker recognition can be completed by recognizing the unique timbre (such as the frequency) of the speaker, which is unrelated to the language they speak. Thus, both acoustic scene classification and speaker recognition can be language-agnostic tasks.}
	
	{In contrast, we generally realize someone has a certain accent when hearing that a specific pronunciation is different from the standard one. Therefore, for the AR task, to judge whether someone has a different accent, the knowledge of that language is needed. The discriminative feature of the accented pronunciation is also more subtle. As a result, we think that AR differs from acoustic scene classification or speaker recognition because AR requires a more fine-grained and language-related feature, which is more difficult compared to the language-agnostic tasks.}  
	
	In this paper, firstly, we {review the related AR papers and systems} in Sec.\ref{sec:related}. 
	We give the proposed method a detailed description in Sec.\ref{sec:method}. We use a CNN-based ASR frontend to encode the language-related information. Then we append extra self-attention layers and an aggregation layer to capture the time sequence relevance and conduct classification. We apply the ASR loss as an auxiliary multi-task learning (MTL)\cite{Crawshaw2020911} target to extract  language-related acoustic features. To make the acoustic features more robust, we fuse the embedding of a {trainable} acoustic model that is trained on the accented dataset and a {fixed} acoustic model trained on the standard dataset for a hybrid aggregation.
	
	We conduct dense experiments to compare the model architecture and analyze different training methods in Sec.\ref{sec:exp}. Finally, we give our conclusion in Sec.\ref{sec:con}.
	
	The contributions of this paper are summarized as follows.
	\begin{itemize}
		\item We propose a novel hybrid structure to solve the AR task. Our method adopts an {ASR} backbone and fuses the language-related phonetic features to perform the AR task along with ASR MTL. 
		
		\item We investigate the relationship between ASR and AR. Specifically, we find that without the auxiliary ASR task, AR may easily overfit to the speaker label in the trainset. The proposed ASR MTL alleviates such a phenomenon. Moreover, the {phonetic} acoustic features extracted by the hybrid acoustic models can make the AR model more robust.
		
		\item We conduct dense experiments on Accented English Speech Recognition Challenge (AESRC) 2020 dataset\cite{Shi2021220}. The results show that the recognition accuracy of the proposed method is 6.57\% better than the official baseline on the validation set. We also get a 7.28\% relative improvement on the final test set in the related AESRC competition\footnote{ Competition results are available at
			https://www.datatang.ai/interspeech2020, Track1 or \cite{Shi2021220}. Our team code is Z2 and we rank the 3rd. A detailed description of the dataset is also available in both of them.}.
	\end{itemize}

	\begin{figure*}[h!]
	\centering
	\includegraphics[width=0.8\linewidth]{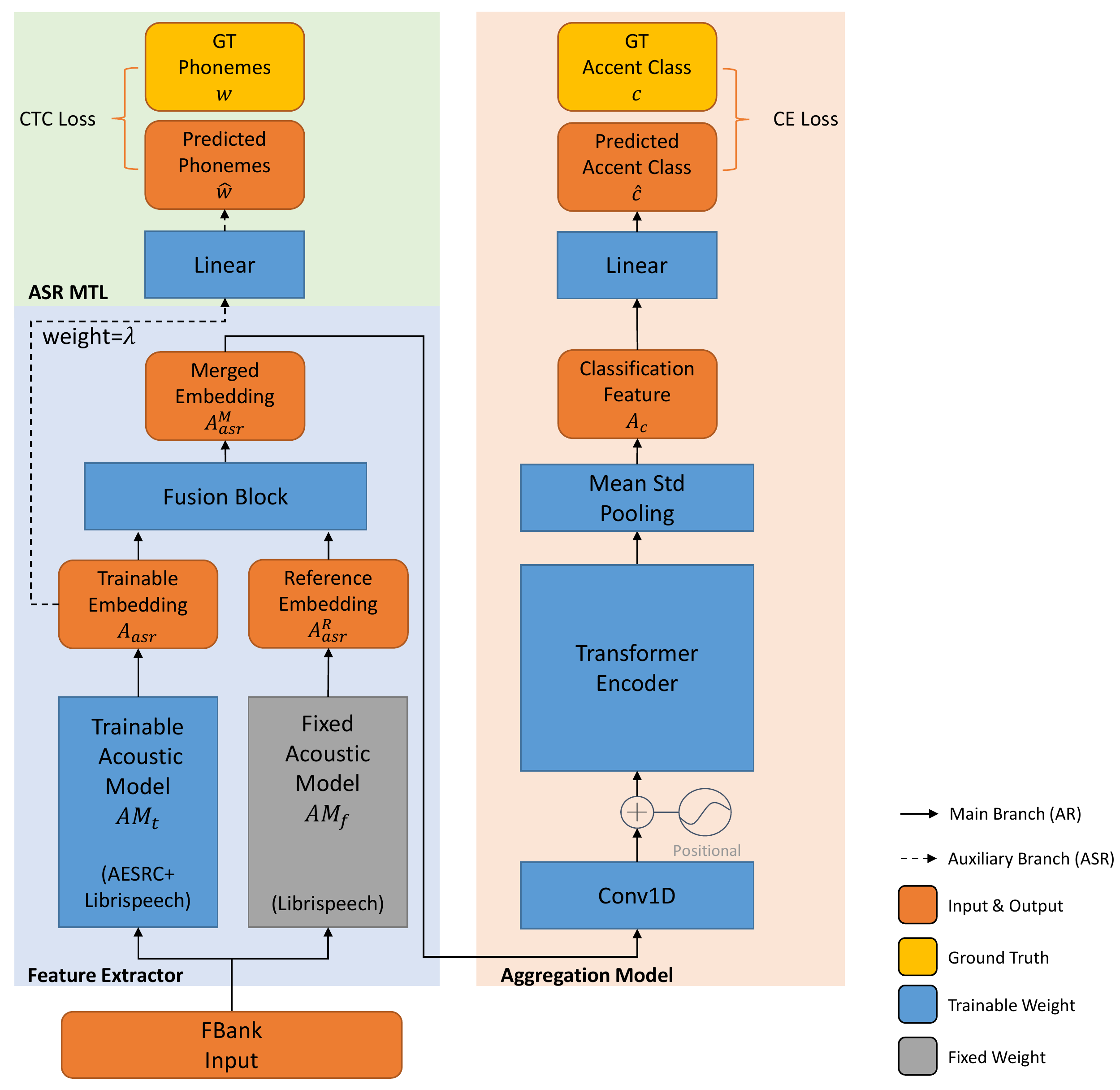}
	\caption{{Proposed hybrid structure for accent recognition. We list the dataset used for the acoustic model ($AM$) in the bracket and use the gray color to indicate the fixed acoustic model does not participant in the AR training process. The auxiliary branch plotted in dash line (the green block) is used only during training.}}
	\label{fig:model}
\end{figure*}

\begin{figure*}[!h]
	\centering
	\begin{minipage}[b]{0.3\linewidth}
		\includegraphics[width=1\linewidth]{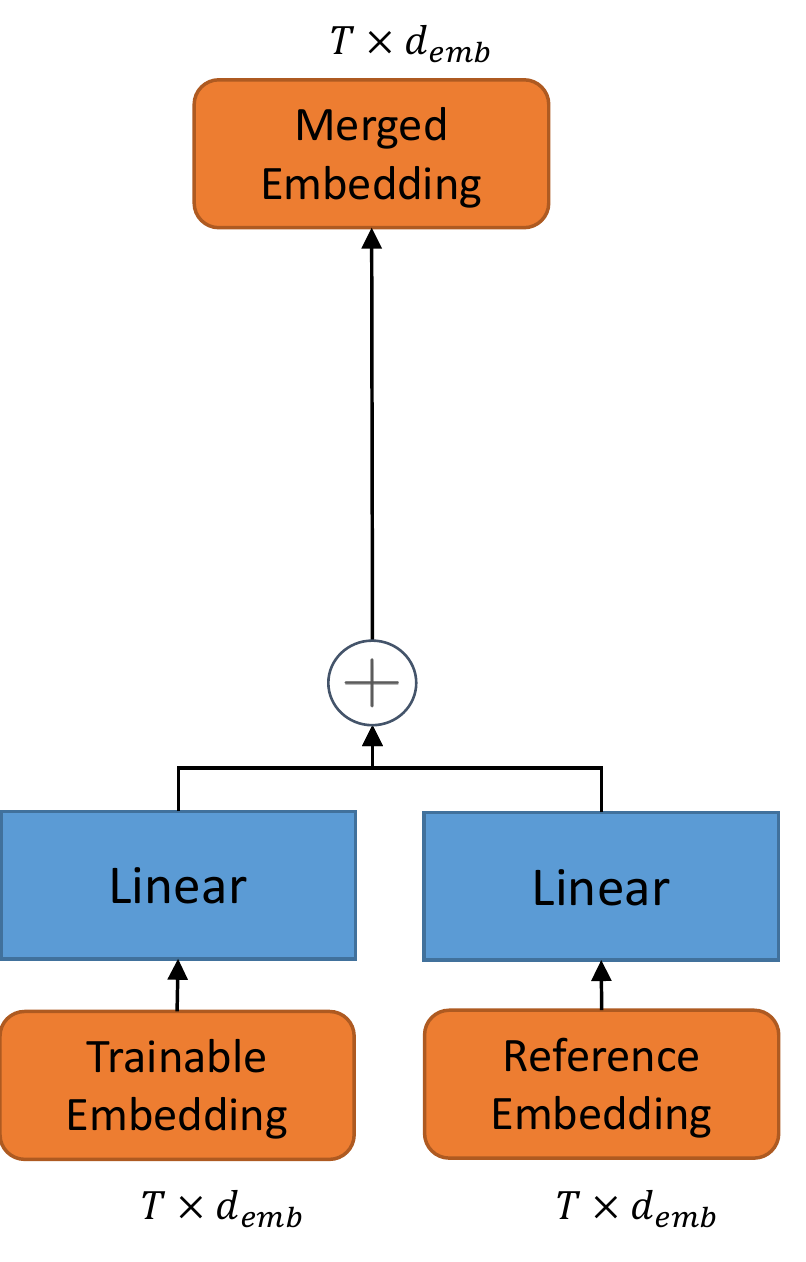}
		\subcaption{}
		\label{fig:add_fusion}
	\end{minipage}
	\begin{minipage}[b]{0.3\linewidth}
		\includegraphics[width=1\linewidth]{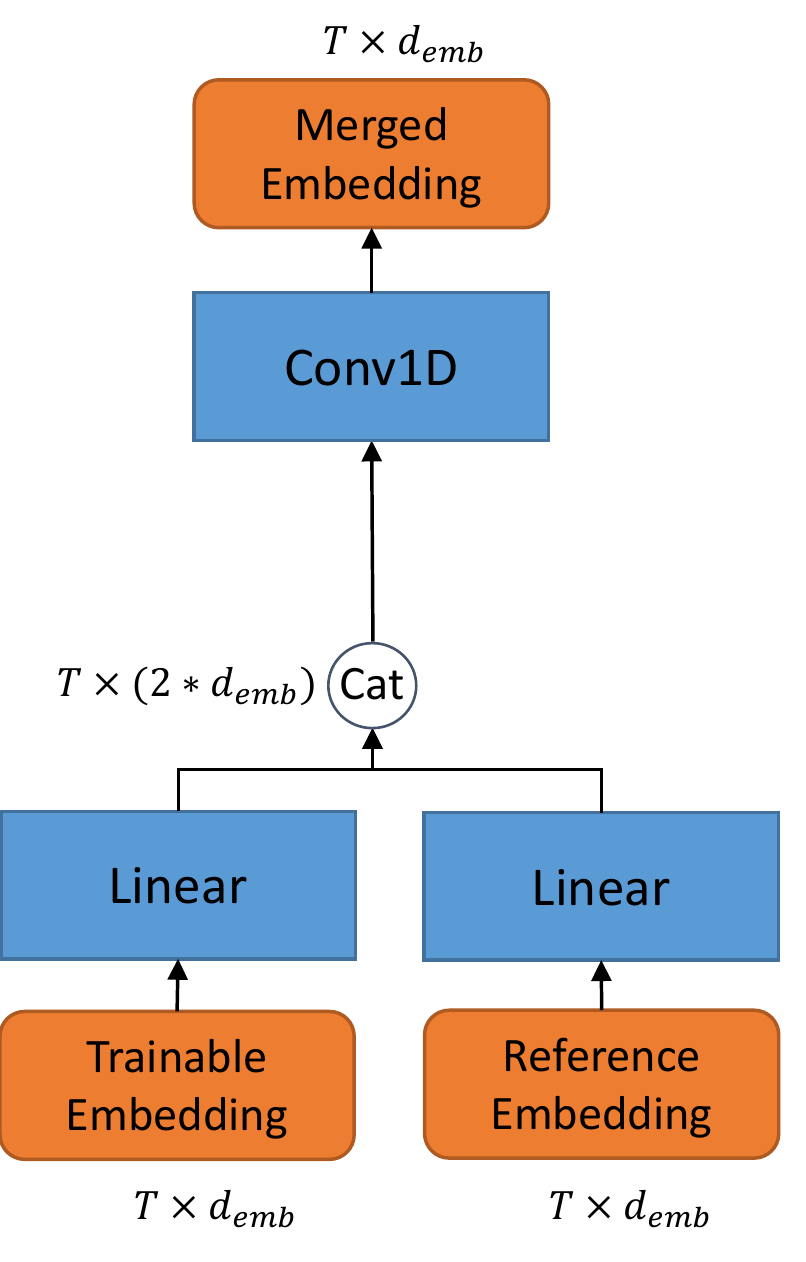}
		\subcaption{}
		\label{fig:cat_fusion}
	\end{minipage}
	\begin{minipage}[b]{0.3\linewidth}
		\includegraphics[width=1\linewidth]{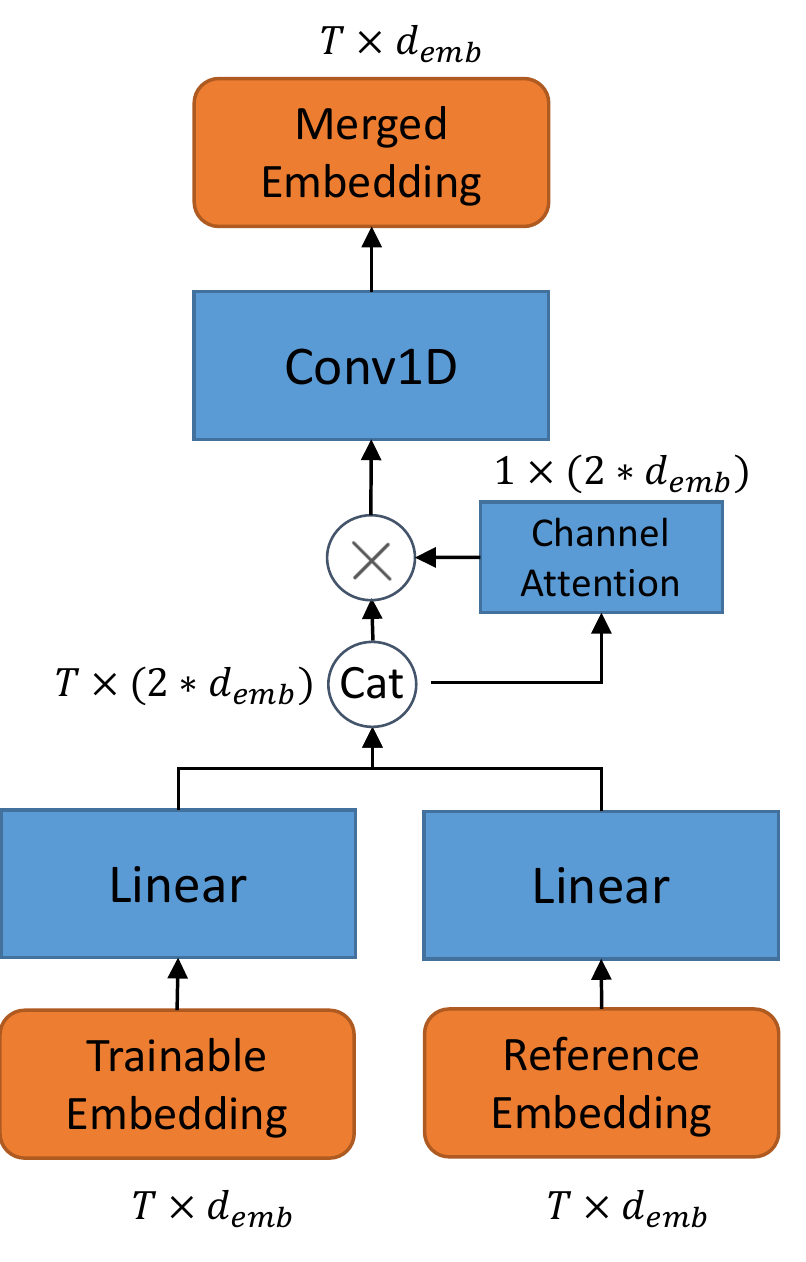}
		\subcaption{}
		\label{fig:cat_ca_fusion}
	\end{minipage}	
	\caption{{Fusion block with different structures.} (a) Adding-based fusion. (b) Concatenation-based fusion. (c) Concatenation-ChannelAttention-based fusion. In (b)(c), a Conv1D layer with kernelsize=1 is adopted to cast the concatenated channels $2*d_{emb}$ back to $d_{emb}$.}
	\label{fig:fusion}
	
\end{figure*}

\begin{table*}[!h]	
	\centering
	\begin{tabular}{lll}
		\textbf{Module} & \textbf{Output Shape} & \textbf{Output Description} \\ \hline
		FBank & $T_{in}\times d_{fbank}$ & Input feature, $d_{fbank}=40$ \\
		$AM_t$ & $T\times d_{emb}$ & Trainable embedding, $A_{asr}$, $d_{emb}=1024$ \\
		Linear & $T\times d_{phoneme}$ & Phoneme prediction, $\hat{w}$, $d_{phoneme}=40$ \\
		$AM_f$ & $T\times d_{emb}$ & Reference embedding, $A_{asr}^{R}$\\
		Fusion Block & $T\times d_{emb}$ & Merged embedding, $A_{asr}^M$ \\
		Conv1D(kernelsize=1) & $T\times d_{attn}$ & Reshaped embedding for aggregation, $d_{attn}=256$ \\
		Self-Attention Aggregation & $1\times d_{attn}$ & Aggregated feature for classification, $A_c$ \\
		Linear & $1\times d_{accent}$ & Predicted accent class, $\hat{c}$, $d_{accent}=8$
		
	\end{tabular}
	\caption{Output summary of the proposed model. $T_{in}$ is the length of the input sequence. Note that the acoustic model will downsample the feature sequence by 2 on the time domain, i.e., $T_{in}=2T$.}
	\label{tb:io}
\end{table*}	
	
	\section{Related Work}
	\label{sec:related}
	{In this section, we start from  reviewing the related AR papers. Next, we analyse the other systems submitted to the AESRC 2020 evaluation.}

	\subsection{Accent Recognition}
	
	To solve the AR task, conventional methods including  \cite{Mannepalli2016, Behravan2016} adopt Hidden Markov Model (HMM) to model the time features. \cite{Siniscalchi2014} integrates both the speech attribute features and the acoustic cues to better capture foreign accents. \cite{Chen2018} applies Fisher Vector (FV) and Vector of Locally Aggregated Descriptors (VLAD) methods for the language identification task. \cite{Cai2018b} gives a thorough study about the encoding methods and loss functions for the language recognition task.

	Besides the aforementioned works that focus on the recognition task alone, there are other works that apply AR as an auxiliary task to boost the performance of the other main tasks. 
	\cite{Najafian2020, Weninger2019} explore the relationship between accent and the ASR task, and show that the accent recognition task can lead to a more robust performance of the main ASR task.
	
	Although research in this area has shown that AR can be an auxiliary task to boost the ASR performance, when ASR performs as the auxiliary task for AR, the effects have not been closely examined. How to combine AR with the ASR task for a better language-related feature still remains to be investigated. This paucity inspires us to design our novel hybrid AR model combined with ASR MTL.
	
	\subsection{{AESRC Systems}}
	
	As summarized in \cite{Shi2021220}, Transformer\cite{Vaswani2017613}  is the mainstream backbone for the systems submitted to AESRC.  There are also teams that use ResNet\cite{He20151211} or TDNN\cite{Lang1990} to model the accent. Besides the deep learning methods, a combination of neural network and support vector machine (SVM)\cite{Cristianini2000} is also applied for AR by a team. 
	
	As reported, data augmentation plays an important role in the AR training process. SpecAug\cite{Park1519September2019} is used by many teams. In addition, pitch shift, noise augment, and speed perturb are also applied. It is worth noticing that the 1st team uses the provided data to train an accented text-to-speech (TTS) system\cite{Huang2021219}. They synthesize extra accented training audio clips with this TTS system, and the accuracy of AR is improved by 10\% absolutely. 
	
	The submitted systems also show that the language-related information is of vital importance to train the AR task.
	For example, the 1st team uses phone posteriorgrams (PPG) features generated by a Kaldi\cite{article} ASR system as the model input. The 2nd team concatenates the accent label with the text label sequence for a sequence-to-sequence training.
	
	Among these teams, our system ranked the 3rd. We also adopt a phoneme classification task to extract the language-related information. Besides, we pay more attention to the relationship between AR and ASR. We propose a novel hybrid structure to fuse the phonetic features for AR, which will be discussed in the next section.

	\section{Proposed Method}
	\label{sec:method}
	In this section, we first start from a baseline that directly models the relationship between the input spectrum and the accent label. 
	However, such a baseline does not make full use of the spoken words. We analyze that making use of the text information can help the accent modeling in the next subsection. We further propose our ASR MTL method which can incorporate the extra text information. Finally, we analyze the ASR learning target and propose to use hybrid acoustic models for a more robust prediction. The whole model structure is shown in Fig.\ref{fig:model}.
	
	\subsection{Network Structure}
	\label{sec:baseline}
	To summarize, the acoustic scene classification, speaker recognition, and AR tasks discussed in Sec.\ref{sec:intro} all process a sequence into a certain label. As the input feature is generally spectrums, which can be viewed as images, intuitively, we may consider using computer vision (CV) models including VGG\cite{Simonyan201495} or ResNet\cite{He20151211} for classification.  However, as the AR task needs more fine-grained features, we choose to adopt an {ASR-based backbone} to extract the high-level features for each frame, and then aggregate them for the final accent prediction.
	
	
	The attention mechanism can be a good choice to aggregate the extracted acoustic features, and there are also attention-based aggregation methods successfully applied to the accent classification task, including \cite{Chen2018} and the self-attention pooling layer \cite{Cai2018b}.
	However, a single self-attention pooling layer appended to the acoustic model may not be enough. As we also perform an auxiliary ASR task (discussed in the next subsection) for the acoustic model,  the acoustic model still needs to adapt from the ASR task to the AR task. In other words, a single pooling layer may be too shallow to transfer the task. 
	
	For our method, we tend to keep the aggregation model independent from the acoustic model so that the ASR loss applied to the acoustic model can be more efficient. Thus, we choose to stack the multi-head self-attention layers for a high-level extraction of the accent attribute for each frame. This extractor is, in fact, the encoder part of Transformer\cite{Vaswani2017613}. 
	
	Formally, the attention mechanism can be described as,
	\begin{equation}
		Attention(Q, K, V)=softmax\left(\frac{Q K^{T}}{\sqrt{d_{k}}}\right) V,
	\end{equation}
	where $softmax\left({\frac{QK^T}{\sqrt{d_k}}}\right)$ is the dot-product similarity of the query matrix $Q$ and key matrix $K^T$ (scaled by $\frac{1}{\sqrt{d_k}}$), constructing the weight of the value matrix $V$.
	The multi head attention (MHA) is described as,
	\begin{equation}
		MHA(Q,K,V)=concat(hd_1,...,hd_h)W^O,
	\end{equation}
	where
	$hd_i=Attention(QW_i^Q,KW_i^K,VW_i^V)$. $h$ is the head number, and $W^Q$, $W^K$, $W^V$, $W^O$ is the projections matrices for query, key, value, and output, correspondingly. For the self-attention case, $Q=K=V=x$, where $x$ is the input features for each attention layer.
	
	Compared with the self-attention pooling layer, multiple attention heads help the model to capture different representation subspaces at different positions of the accent.
	
	After the stacked multi-head self-attention layers, we use a statistic pooling layer\cite{Snyder2018} to compute the mean and standard deviation across the time dimension of the acoustic sequence. Finally, we use a fully-connected layer to project the aggregated features into the final accent prediction.
	
	In this paper, we use a CNN-based acoustic model called Jasper\cite{Li201946} to extract deep phonetic features from the input 40-dim FBank spectrum. {Jasper is constructed by multi Jasper blocks, and each Jasper block is further constructed by multi CNN-based subblocks (1D Conv, BatchNorm, ReLU, dropout) with residual connections. We use the Jasper 5x3 version (5 Jasper blocks, each Jasper block has 3 sub-blocks) for our experiments.}
	For the aggregation model, we set the attention dim $d_{attn}=256$, the feed-forward dim $d_{ff}=1024$, the head number $h=4$, and the number of stacked multi-head attention layers $n_{layers}=3$.
	
	\subsection{Multi Task Learning}
	\label{sec:mtl}
	As discussed in the aforementioned Sec.\ref{sec:intro}, AR requires a more fine-grained feature compared with acoustic scene classification and speaker recognition. Meanwhile, acoustic scene classification and speaker recognition can be a language-agnostic task, which is easier compared to the language-related AR task. Consequently, the AR task may degrade to the speaker recognition task if we do not set extra {constrains to the learning targets}.
	
	Formally, if we denote the model parameters as $\theta$, the acoustic features for classification as $A_c$, the accent label as $c$, the predicted accent class $\hat{c}$ can be described as,
	\begin{equation}
		\hat{c}=argmax_c P(c|A_c;\theta),c\in\{c_i,...,c_n\},
	\end{equation}
	where $n$ is the number of accents in the dataset.
	
	We use the cross-entropy (CE) loss as the classification loss function between the predicted accent class  $\hat{c}$ and the ground truth label $c$,
	\begin{equation}
		l_c=CE(c,\hat{c}).
	\end{equation}
	
	However, we should note that for the training dataset, the accent attribute is labelled for each speaker. In other words, each accent label $c_j$ can be inferred from the speaker label $s_i$,
	\begin{equation}
		c_j=Accent(s_i),1\leq i \leq m,1 \leq j \leq n,
		\label{eq:accent}
	\end{equation}
	where $m$ is the total number of speakers in the training dataset.
	
	As a result, if we only apply the cross-entropy loss for the accent classification, a possible solution for the network is to simply memorizes each speaker $s$ in the training dataset and learn the mapping between the predicted speaker $\hat{s}'$ and the accent label $\hat{c}'$ by using the following Eq.\ref{eq:sr} and Eq.\ref{eq:accent}:
	\begin{equation}
		\hat{s}'=argmax_s P(s|A_c;\theta),s\in\{s_i,...,s_m\}.
		\label{eq:sr}
	\end{equation}

	In other words, the network will overfit to the speaker label on the trainset. {For unseen speakers, the AR performance will be severely degraded.}  We demonstrate this phenomenon in Sec.\ref{sec:sr_test}. 
	
	{We assume that accent recognition using speaker-invariant features is more accurate than the method via speaker recognition, and we force the network to learn the language-related information.  On the one hand, the text transcription is independent of the speaker	information, and ASR MTL is suitable for this task. On the other hand, ASR MTL can offer complementary information for the main AR task. The ASR task forces the model to learn the language-related information, and the fine-grained phonetic feature can contribute to the accent recognition.}
	We build another branch of the proposed model in Fig.\ref{fig:model} to perform the ASR task {during training}. The auxiliary ASR branch tries to predict the pronounced phonemes  $\hat{w}$ based on the ASR acoustic features $A_{asr}$ and the parameters of the ASR acoustic model $\theta_{asr}$,
	\begin{equation}
		\hat{w}=argmax_w P(w|A_{asr};{\theta}_{asr}).
	\end{equation}
	We convert the text from word-level to phoneme-level using a {grapheme to phoneme (G2P) tool\footnote{https://github.com/Kyubyong/g2p}} and use CTC\cite{Graves2006} as the ASR loss function,
	\begin{equation}
		l_{asr}=CTC(w, \hat{w}).
	\end{equation}
	
	For the proposed model, the feature for accent classification, $A_c$, is aggregated from the feature for ASR, $A_{asr}$,
	\begin{equation}
		A_c=AggregationModel(A_{asr})
		\label{eq:aggregate}.
	\end{equation}
 We use the parameter $\lambda$ to balance the weight between the ASR task and the AR task,	
	\begin{equation}
		l=l_c+\lambda l_{asr}.
		\label{eq:mtl}
	\end{equation}
	
	\subsection{Hybrid {Phonetic Features}}
	\label{sec:hybrid}
	Although the AR performance can be boosted by the auxiliary ASR task, we should note that there is a difference between the ASR target and the AR target. 
	
	Given a certain speech, ASR allows mispronunciations caused by accents and corrects them to the target text. For example, non-native speakers with different accents may pronounce the target word ``AE P AH L'' (APPLE) to ``AE P OW L'' or ``AE P AO L''. We denote this pronounced phoneme as $AH_{OW}$ or $AH_{AO}$. There is no need for ASR models to distinguish from different accents, and ASR models will ignore different accents and still map both $AH_{OW}$ and $AH_{AO}$ accented speech into the non-accented transcription of this dataset, $AH$. In other words, ASR models will not work hard to explore the differences between different accents. 
	
	On the contrary, AR needs to find the differences between the 
	 $AH_{OW}$ and $AH_{AO}$ referring to the original $AH$. 
	This conflict inspired us to use another acoustic model trained on the non-accented dataset for a fixed reference.
	
	For the proposed method, {two Jasper acoustic models are used. The first one is trained on the non-accented dataset (Librispeech). The second one is first trained on the non-accented dataset and then trained on the accented dataset (AESRC, these two datasets will be introduced later). We pretrain these two acoustic models with the CTC loss and keep the model with the lowest validation phone error rate (PER) for further experiments. We freeze the weight parameters of the non-accented acoustic model (AM) and call it the ``fixed AM'' (denoted as $AM_f$). The accented one is further fine-tuned together with the aggregation model using Eq.\ref{eq:mtl}, and we call it the ``trainable AM'' (denoted as $AM_t$).}  As illustrated in Fig.\ref{fig:model}, we merge the reference phonetic embedding of the fixed AM into the trainable one with a fusion block.
	
	{In this paper, we consider three different fusion blocks as illustrated in Fig.\ref{fig:fusion}.}
	First, a linear projection is applied to the embeddings,
	\begin{equation}
		A_{asr}^{'}=linear\left(A_{asr}\right),
	\end{equation}
	\begin{equation}
	A_{asr}^{R'}=linear\left(A_{asr}^{R}\right).
\end{equation}	

	\textbf{Adding-based fusion}. As illustrate in Fig.\ref{fig:add_fusion}, an add function is applied to merge the processed embeddings.
	\begin{equation}
		A_{asr}^{M}=A_{asr}^{'}+A_{asr}^{R'}.
	\end{equation}
	However, simply adding them may lose the information of the raw data. 
	
	\textbf{Concatenation-based fusion.}
	Instead, we stack the embeddings on the channel domain and use a Conv1D(kernelsize=1) to half the channels of the concatenated embedding $A_{asr}^C$. This structure is illustrated in Fig.\ref{fig:cat_fusion}. This process can be described as,
	\begin{equation}
		A_{asr}^{C}=ChannelCat\left(A_{asr}^{'}, A_{asr}^{R'}\right),
		\label{eq:cat}
	\end{equation}
	\begin{equation}				A_{asr}^{M}=Conv\left(A_{asr}^{C}\right),
	\end{equation}
	where $A_{asr}^{C}\in\mathbb{R}^{T\times 2d_{emb}}$ and $A_{asr}^{M}\in\mathbb{R}^{T\times d_{emb}}$.
	
	\textbf{Concatenation-ChannelAttention-based fusion.} Furthermore,  we adopt channel-attention to better control the importance of the reference embedding as illustrated in Fig.\ref{fig:cat_ca_fusion}.
	The channel-attention is obtained by the squeeze-excite block\cite{Hu2018}\cite{Woo2018}. Based on Eq.\ref{eq:cat}, the concatenated feature $A_{asr}^C$ goes through a global MaxPooling and a global AveragePooling on the time domain, 
	\begin{equation}
		C=TimeMax(A_{asr}^{C})+
		TimeAve(A_{asr}^{C}),
	\end{equation}
where $C\in\mathbb{R}^{1\times 2d_{emb}}$. $C$ is further squeezed and excited,
\begin{equation}
	   C_{squeeze}=ReLU(linear(C)),
\end{equation}
\begin{equation}
	C_{excite}=ReLU(linear(C_{squeeze})),
\end{equation}
where $C_{squeeze}\in\mathbb{R}^{1\times \frac{2d_{emb}}{r}}$ and $C_{excite}\in\mathbb{R}^{1\times 2d_{emb}}$. We choose $r=16$.

The channel-attention $CA$ is obtained by the sigmoid activation of $C_{excite}$, 
\begin{equation}
	   CA=Sigmoid(C_{excite}),
\end{equation}
and used as the scaling factor of $A_{asr}^{C}$ on the channel domain,
\begin{equation}
	A_{asr}^{M}=Conv\left(CA\cdot A_{asr}^{C}\right).
\end{equation}

	Finally, for these three fusion methods, we aggregate the merged ASR feature $A_{asr}^M$ instead of the original $A_{asr}$ in Eq.\ref{eq:aggregate} and optimize the whole model with the aforementioned MTL loss (Eq.\ref{eq:mtl}) for AR classification.
	We use 39 English phonemes plus $\left\langle \operatorname{BLANK} \right\rangle$  for CTC classification, and there are 8 accents included in the AESRC dataset. The detailed model description is summarized in Tab.\ref{tb:io}.
	
	\begin{figure*}[!h]
	\centering
	\begin{minipage}[b]{0.35\linewidth}
		\centering
		\includegraphics[width=1\linewidth]{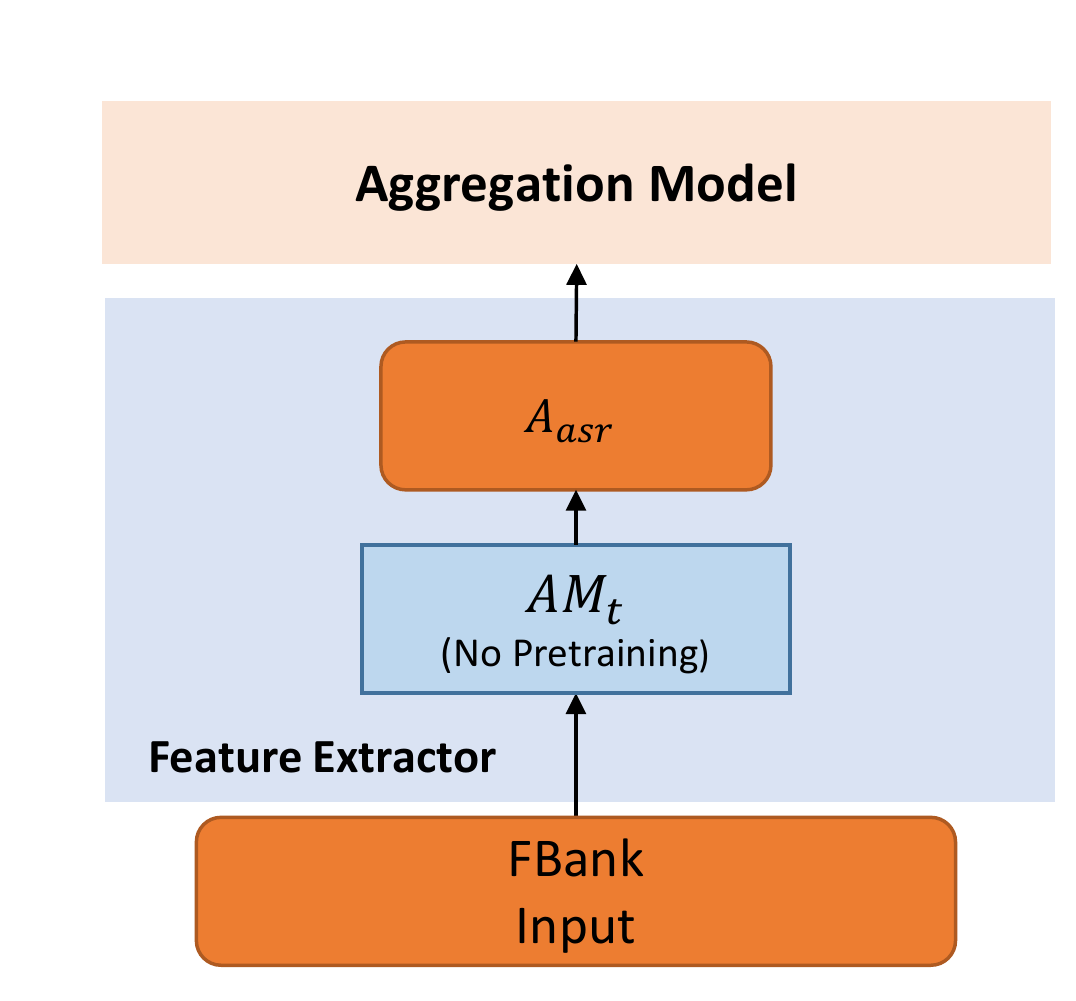}
		\subcaption{}
		\label{fig:baseline}
	\end{minipage}
	\centering
	\begin{minipage}[b]{0.35\linewidth}
		\centering
		\includegraphics[width=1\linewidth]{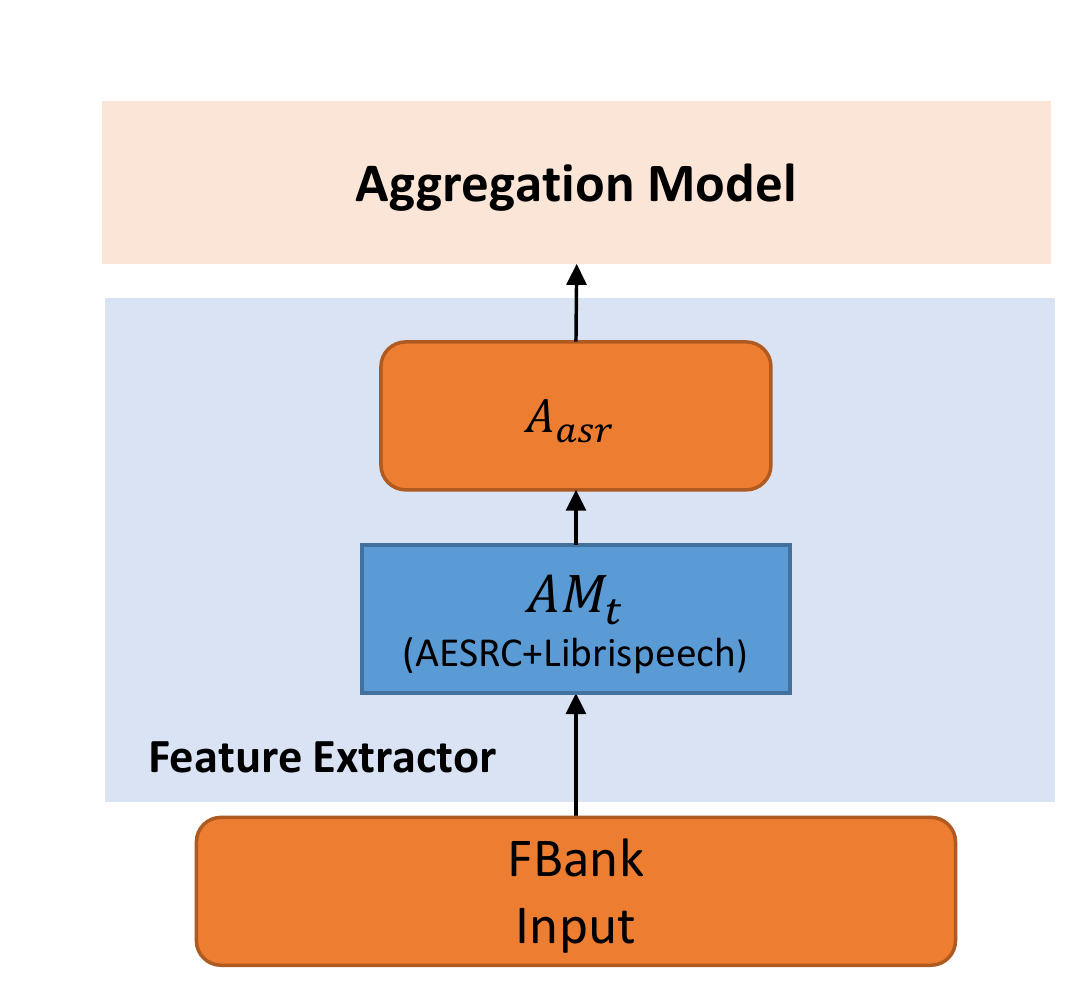}
		\subcaption{}
		\label{fig:asr_init}
	\end{minipage}
	\centering
	\begin{minipage}[b]{0.35\linewidth}
		\centering
		\includegraphics[width=1\linewidth]{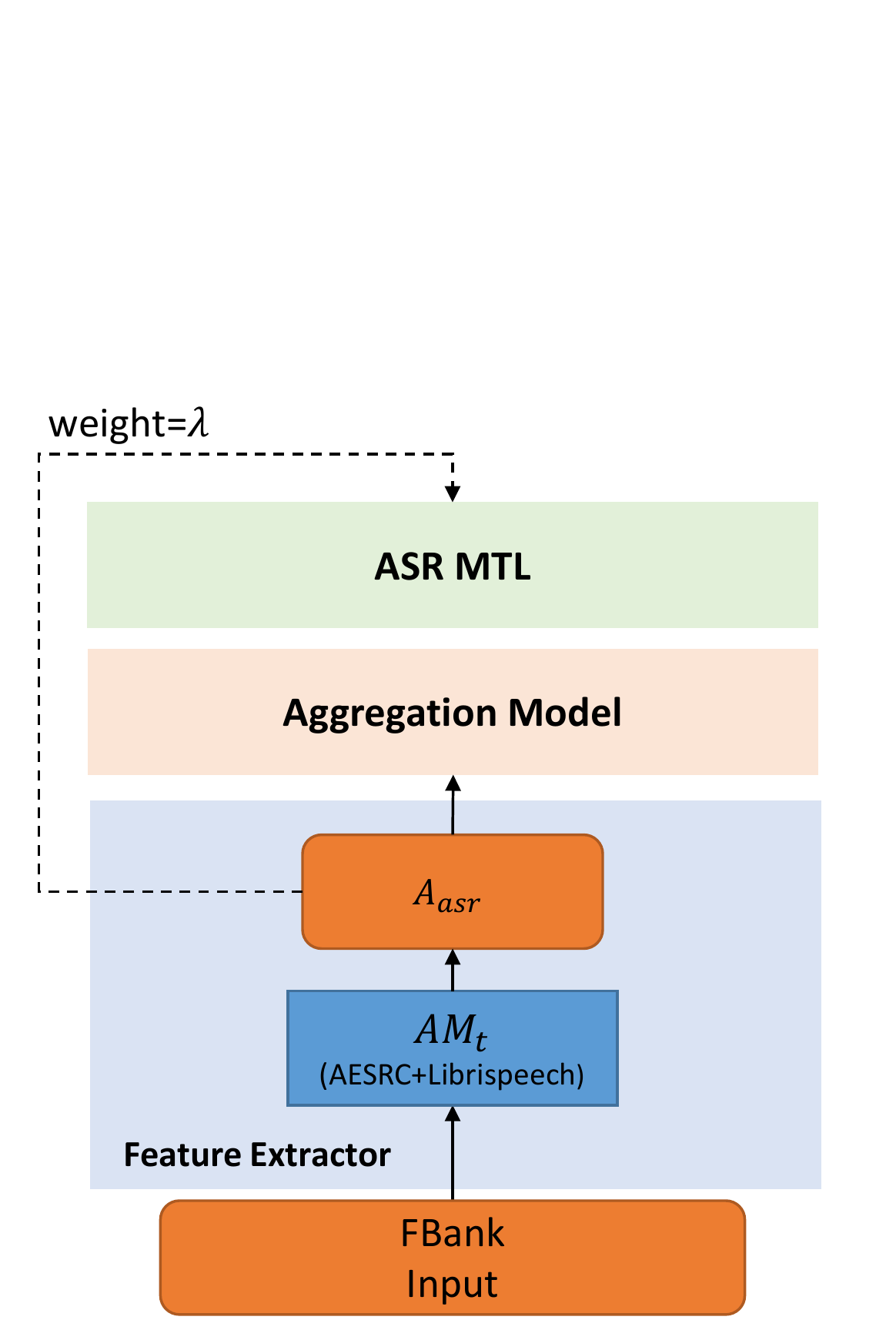}
		\subcaption{}
		\label{fig:mtl}
	\end{minipage}	
	\begin{minipage}[b]{0.35\linewidth}
		\centering
		\includegraphics[width=1\linewidth]{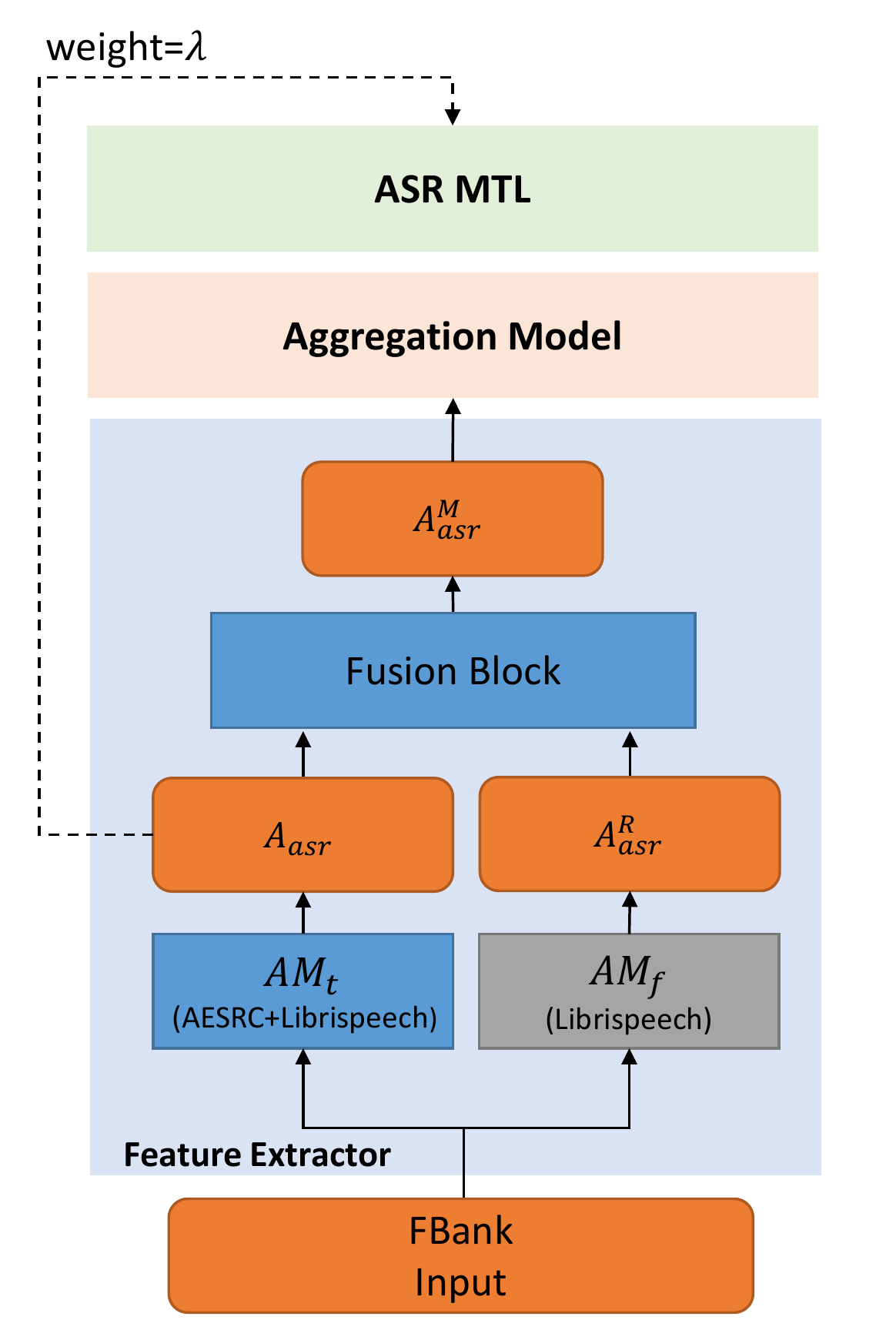}
		\subcaption{}
		\label{fig:hybrid}
	\end{minipage}
	
	\caption{{Different training methods.} (a) Directly train the whole model for the AR task. (b) Pretrain the acoustic model for ASR and then train the whole model for AR. (c) Pretrain the acoustic model for ASR and then train the whole model for AR and ASR MTL (weight=$\lambda$). (d) The proposed hybrid method. Based on (c), the non-accented embedding is merged for reference.}
	\label{fig:compare}
\end{figure*}

	\begin{table*}[!htbp]
	\centering
	\begin{tabular}{lllllllllll}
		\hline
		\textbf{Id} &
		\textbf{Model} & \textbf{US} & \textbf{UK} & \textbf{CN} & \textbf{IN} & \textbf{JP} & \textbf{KR} & \textbf{PT} & \textbf{RU} & \textbf{Total} \\ \hline
		0&Trans 3L           & 45.7 & 70.0 & 56.2 & 83.5 & 48.5 & 45.0 & 57.2 & 30.0 & 54.1 \\
		1&Trans 6L           & 30.6 & 74.5 & 50.9 & 75.2 & 44.0 & 43.7 & 65.7 & 34.0 & 52.2 \\
		2&Trans 12L           & 21.2 & 85.0 & 38.2 & 66.1 & 42.7 & 26.0 & 51.8 & 49.6 & 47.8 \\
		3&Trans 12L(ASR initialized) & 60.2 & 93.9 & 67.0 & 97.0 & 73.2 & 55.6 & 85.5 & 75.7 & 76.1 \\
		4&Ours(No pretraining, Fig.\ref{fig:baseline})      & 45.7 & 81.0 & 40.1 & 79.1 & 45.7 & 37.8 & 84.5 & 63.6 & 60.0 \\
		5&Ours(ASR initialized, Fig.\ref{fig:asr_init})      & 40.3 & 93.7 & 75.0 & 97.3 & 76.3 & 52.1 & 88.3 & 76.0 & 75.1 \\
		6&Ours(MTL, Fig.\ref{fig:mtl}, $\lambda=0.1$)   &  68.6 & 91.8 & 86.9 & 99.1 & 71.2 & 56.6 & 89.5 & 76.0 & 79.9 \\
		7&Ours(MTL, Fig.\ref{fig:mtl}, $\lambda=0.2$) & 57.8 & 92.0 & 79.5 & 98.5 & 70.6 & 71.3 & 84.7 & 66.2 & 77.5 \\
		8&Ours(MTL, Fig.\ref{fig:mtl}, $\lambda=0.3$) & 60.9 & 85.3 & 85.1 & 98.9 & 66.1 & 69.6 & 80.8 & 63.2 & 76.0 \\
		9&Ours(Hybrid, Fig.\ref{fig:hybrid},\ref{fig:add_fusion},$\lambda=0.1$)* & 50.4 & 95.0 & 83.3 & 99.4 & 72.1 & 73.5 & 92.9 & 77.0 & 80.6 \\ 
		10&Ours(Hybrid, Fig.\ref{fig:hybrid},\ref{fig:cat_fusion},$\lambda=0.1$) &61.8 &88.9 &89.6 &98.9 &71.9 &66.0 &95.1 &74.5 &80.8 \\
		11&Ours(Hybrid, Fig.\ref{fig:hybrid},\ref{fig:cat_ca_fusion},$\lambda=0.1$)
		&63.1 &92.3 &88.7 &98.3 &73.9 &66.3 &95.3 &73.7 &81.1 \\
		\hline		
	\end{tabular}
	
	\caption{{Validation accuracy} on AESRC.	(* Our team submitted this adding-based hybrid version for the final test of AESRC.)}
	\label{tab:ar_acc}
\end{table*}

	\section{Experiments}
	\label{sec:exp}
	In this section, we first give a detailed description of the dataset and the experiment environment. We show the AR results in the following subsection.  Next, we conduct a speaker recognition test to demonstrate that the auxiliary ASR task can be of vital importance to keep the AR task from overfitting of the speaker recognition task. Finally, we explore the relationship between the given transcription for the ASR task and the performance of the AR task to test the robustness. 
	
	\subsection{Dataset and Environment}
	We use two specified datasets as required by AESRC for training in our experiments. 
	
	The first one is Librispeech\cite{Panayotov2015}, which does not contain accent labels. This dataset is constructed by approximately 1000 hours of 16kHz read English speech. 
	
	The second one is the officially released dataset by AESRC, which is composed of 8 different accents, including
	Chinese (CN), Indian (IN), Japanese (JP), Korean (KR), American (US), British (UK),
	Portuguese (PT), Russian (RU).
	Each accent is made up of 20 hours of speech data collected from around 60 speakers on average. {The speech recordings are collected in relatively quiet environments with cellphones and provided in Microsoft wav format (16kHz, 16bit and mono).}
	This dataset is split into trainset, valset, and testset for the challenge of AR (track1) and ASR (track2) tasks. 
	Utterances read by certain specific speakers are kept for the valset. 
	
	We use Pytorch\cite{Paszke2019124} to build our systems. We use the 40-dim FBank spectrum features extracted by Kaldi toolkit\cite{article} for the input. Furthermore, we apply SpecAug\cite{Park1519September2019} to perform the data augmentation. We conduct the experiments on a server with Intel Xeon E5-2680 CPU, and 4 NVIDIA P100 GPUs.
	We use the Adam optimizer ($lr=10^{-4}$) for both the ASR pretraining and the AR task. 
	
	\subsection{Accent Recognition Test}
	We show the accent recognition results {on the validation set of} different models in Tab.\ref{tab:ar_acc}. We first start from the official AESRC baseline systems. The official baseline systems use the encoder of Transformer and a statical pooling to perform the AR task. In Tab.\ref{tab:ar_acc}, they are denoted as Transformer-$X$L, where $X$ represents the number of the encoder layers. As we can see from Tab.\ref{tab:ar_acc}, a bigger model {(from Id 1 to Id 3)} will lead to overfitting and a degraded classification accuracy. However, as we can observe from both the baseline models and our models, this overfitting phenomenon can be alleviated by the ASR task. {ASR pretraining on Librispeech and AESRC dataset (Id 3, Id 5)} can greatly improve the AR accuracy compared with the model without pretraining {(Id 2, Id 4, correspondingly)}.  Furthermore, the proposed MTL  version {(Id 6-8)} is better than ASR initiation only. We obtain the best result of the MTL-based models by setting the ASR weight to $\lambda=0.1$. By merging the outputs of the fixed AM and the trainable AM, the proposed hybrid methods {(Id 9-11)} show a better performance compared with the MTL version. {The Concatenation-ChannelAttention-based fusion method (Id 11)} shows the best performance among these models, which results in a 6.57\% relative improvement over the best official ASR initialized baseline {(Id 3)}. {Note that the performance of US data is relative low. We think this is because the accented dataset supposes that the data collected in each country belongs to one type of accent. Thus, the accent label is decided by which country the speaker belongs to, rather than their native language. As speakers in the US are various, the performance
	for US can be downgraded.}

\begin{table*}[!h]
	\centering
	\begin{tabular}{llllllllllll}
		\textbf{Transcription} & \textbf{Id}& \textbf{Model} & \textbf{US} & \textbf{UK} & \textbf{CN} & \textbf{IN} & \textbf{JP} & \textbf{KR} & \textbf{PT} & \textbf{RU} & \textbf{Total} \\ \hline
		\multirow{2}{*}{0\%   Degraded($\theta=0$)}    &12& MTL    & 68.6 & 91.8 & 86.9 & 99.1 & 71.2 & 56.6 & 89.5 & 76.0 & 79.9 \\
		&13& Hybrid &63.1 &92.3 &88.7 &98.3 &73.9 &66.3 &95.3 &73.7 &81.1 \\
		\multirow{2}{*}{50\%   Degraded($\theta=0.5$)} &14& MTL    & 54.5 & 93.7 & 91.9 & 99.6 & 62.6 & 67.2 & 79.6 & 66.2 & 76.8 \\
		&15& Hybrid & 50.1 &91.8 &88.3 &98.6 &72.9 &63.7 &94.6 &73.3 &79.3 \\
		\multirow{2}{*}{100\%   Degraded($\theta=1$)}  
		&16& MTL    & 49.0 & 93.6 & 86.6 & 99.5 & 56.6 & 58.4 & 76.7 & 77.1 & 74.7 \\
		&17& Hybrid & 52.6 &90.9 &78.6 &98.3 &72.7 &59.6 &97.5 &68.9 &77.5\\
		\multirow{2}{*}{Random}  
		&18& MTL    & 28.8 &81.9 &33.6 &73.3 &39.1 &34.9 &73.3 &45.1 &51.5 \\
		&19& Hybrid &46.6 &73.5 &61.8 &88.6 &49.7 &43.0 &86.6 &68.1 &64.8\\
		\hline
	\end{tabular}
	\caption{{Validation accuracy for} the robustness test under different transcription situations. All models use $\lambda=0.1$. For the hybrid version, we use the channel-attention based fusion (Fig.\ref{fig:cat_ca_fusion}). Note that Id 12, 13 is in fact the same as Id 6, 11 in Tab.\ref{tab:ar_acc}, correspondingly.}
	\label{tab:rob}
\end{table*}
	
	\begin{figure*}[h!]
	\centering
	\begin{minipage}[b]{0.4\linewidth}
		\centering
		\includegraphics[width=1\linewidth]{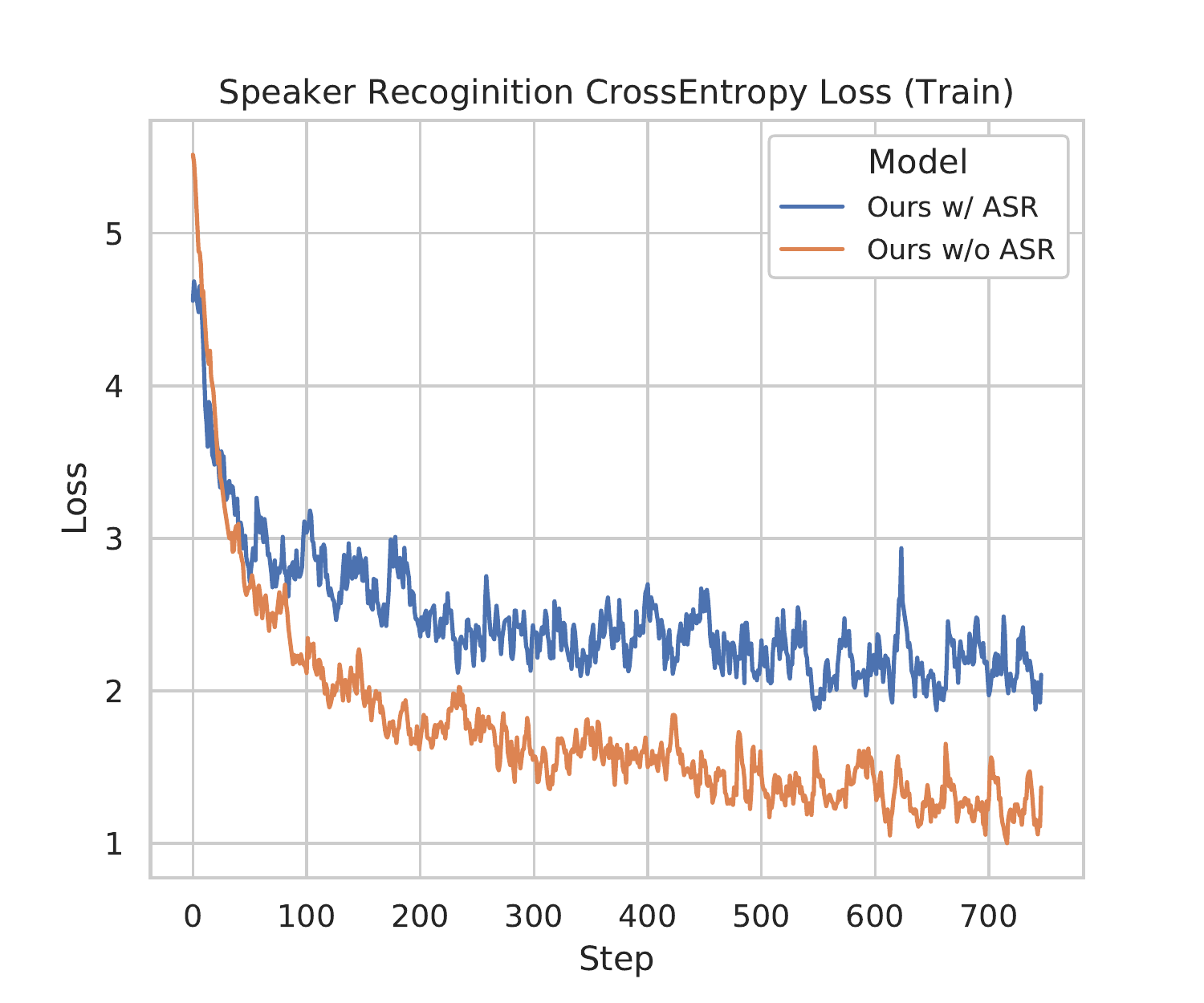}
		\subcaption{}
		\label{fig:cespeaker}
	\end{minipage}
	\centering
	\begin{minipage}[b]{0.4\linewidth}
		\centering
		\includegraphics[width=1\linewidth]{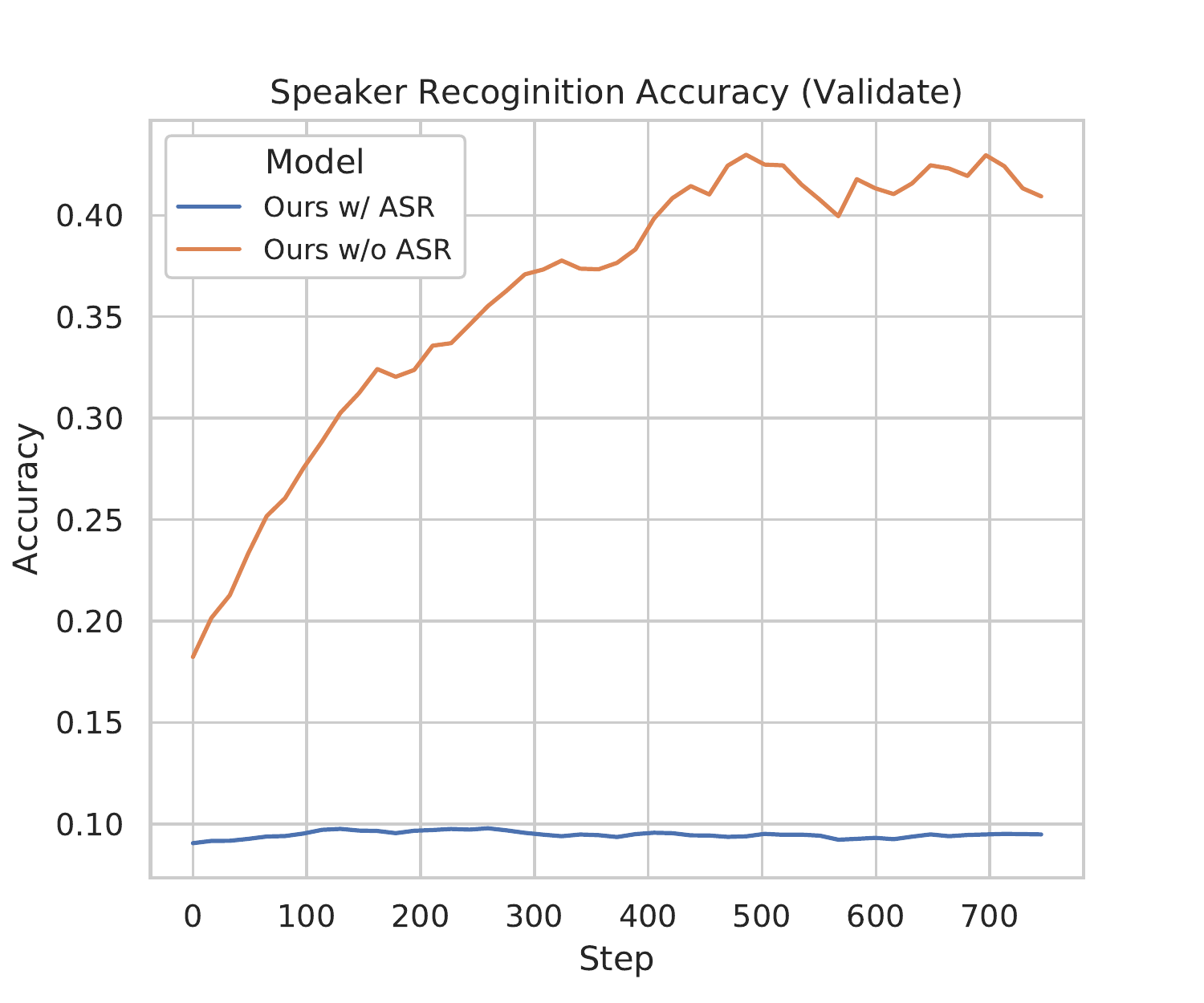}
		\subcaption{}
		\label{fig:accspeaker}
	\end{minipage}
	\caption{Speaker recognition test. (a) Cross-entropy  loss curve for the speaker label while training the models with/without the ASR task. (b) Speaker recognition accuracy curve on the validation dataset while training the models with/without the ASR task.}
\end{figure*}

\begin{figure*}[!h]
	\centering
	\begin{minipage}[h]{0.4\linewidth}
		\includegraphics[width=\linewidth]{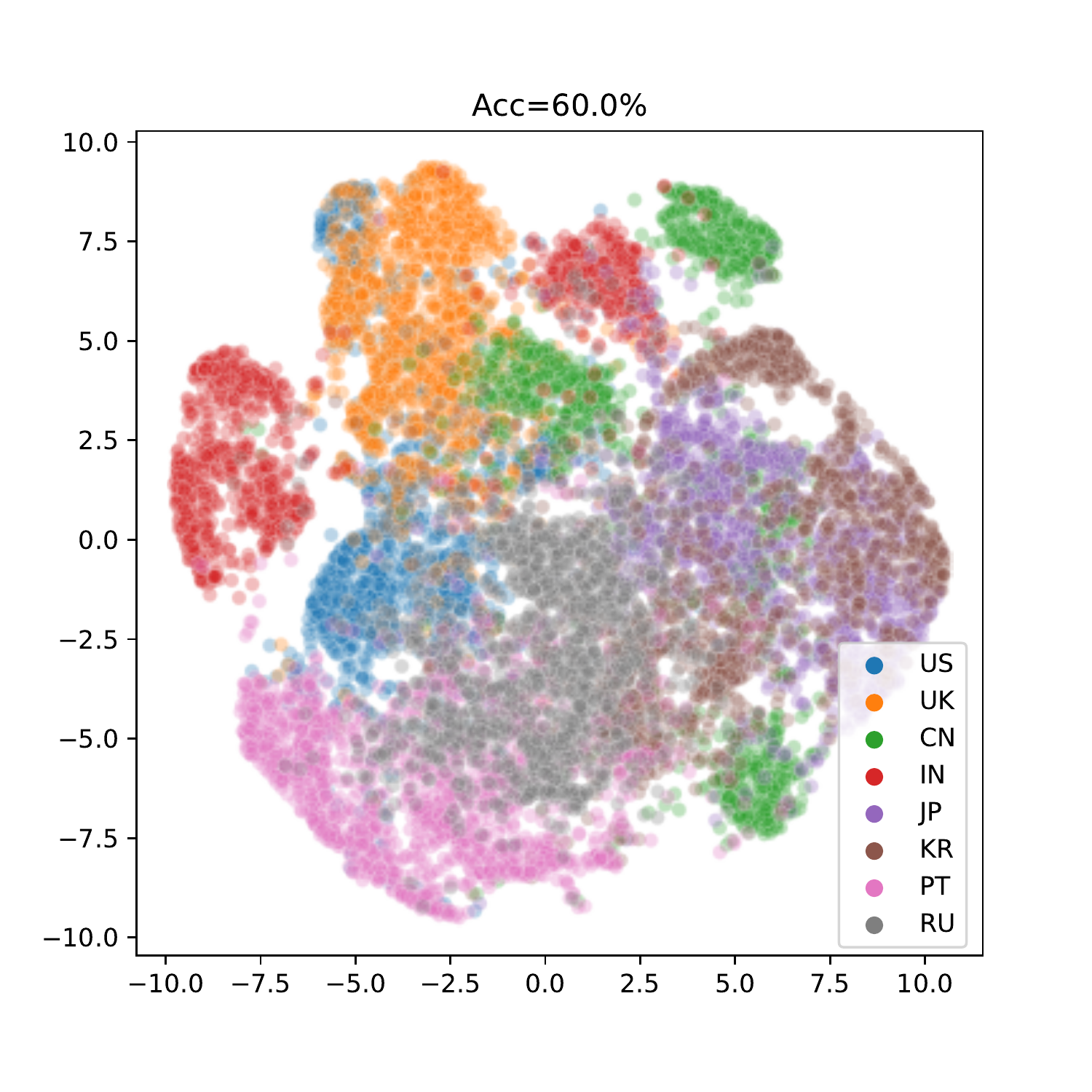}
		\subcaption{}
		\label{fig:e_baseline}
	\end{minipage}
	\begin{minipage}[h]{0.4\linewidth}
		\includegraphics[width=\linewidth]{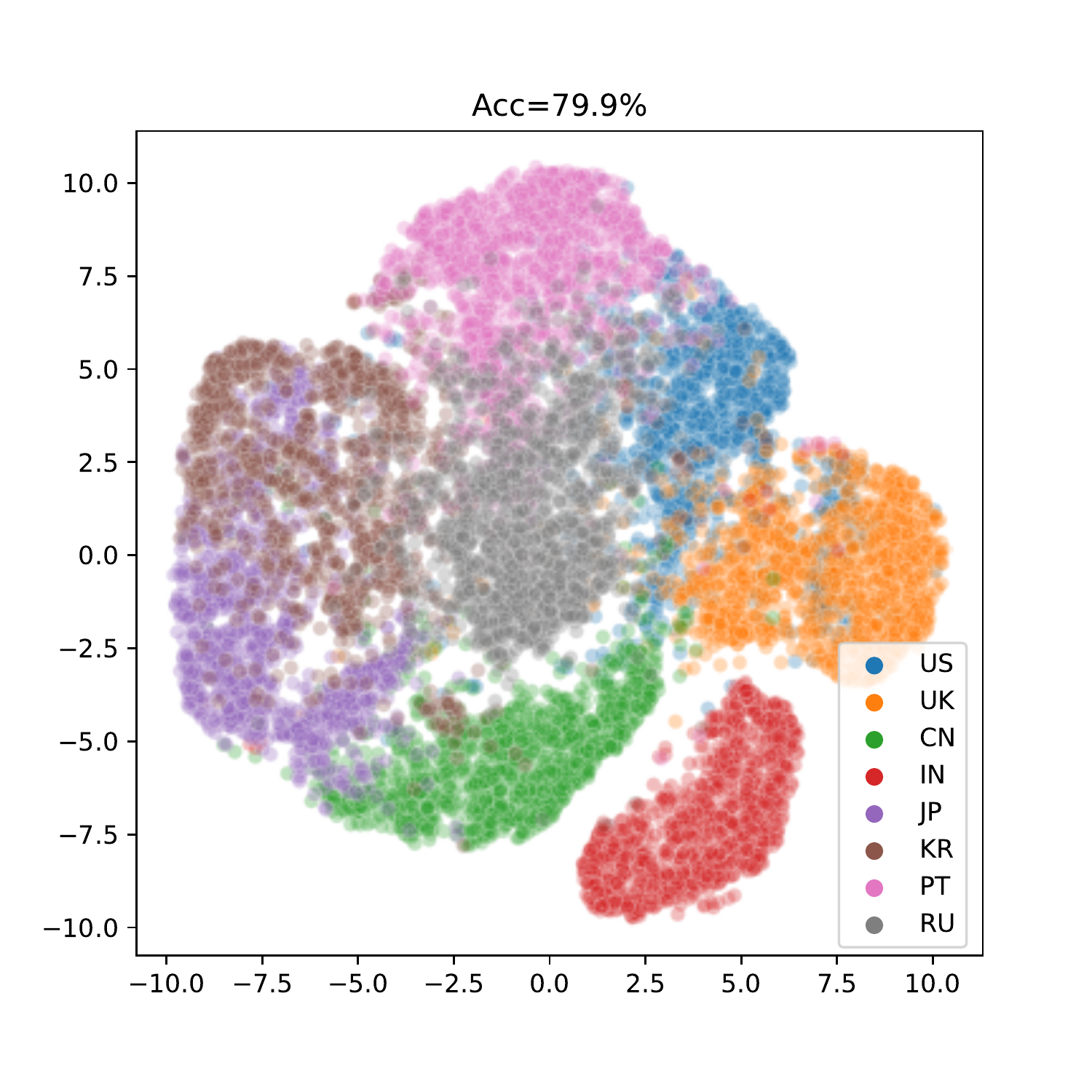}
		\subcaption{}
		\label{fig:e_proposed}
	\end{minipage}	
	\caption{2D embeddings of the accent features. (a) The baseline version without the auxiliary ASR task. (b) The MTL version.}
	\label{fig:emb}
	
\end{figure*}	

	\subsection{Speaker Recognition Test}
	\label{sec:sr_test}
	
	{As discussed above, the model with ASR MTL is better than the one without the ASR pretraining (Id 6, Id 4 in Tab.\ref{tab:ar_acc})}. To validate our assumption in Sec.\ref{sec:mtl}, {that directly training the network for AR may overfit to the speaker label on the trainset, and ASR MTL can help the model to learn a speaker-invariant feature thus alleviating this phenomenon, we compare the speaker recognition performances of these two models.}
	
	{To probe whether the output of these two models contains speaker information,} we freeze the weight parameters of these two models and replace the final linear layer of the aggregation model to perform the speaker recognition task. We only train this linear layer to see how the original feature extracted by each model correlates with the speaker information. We train the linear layer with the cross-entropy loss and Adam optimizer ($lr=10^{-4}$). We should note that for the original AR experiments, the training and validation dataset is split by different speakers. To perform the speaker recognition task, we did another splitting by utterances, and a certain speaker appears in both the training set and the validation set.
	
	We show the training process in Fig.\ref{fig:cespeaker} and the validation process in Fig.\ref{fig:accspeaker}. The model without ASR pretraining has a lower training cross-entropy loss for the speaker label and a much higher speaker recognition accuracy compared to the ASR MTL version. {Such a phenomenon suggests that adding the ASR task indeed helps the model to learn a speaker-invariant feature, and this feature is more powerful to solve the AR task as shown in Tab.\ref{tab:ar_acc}. We also plot the embeddings of the predictions on the validation dataset in Fig.\ref{fig:emb} using t-SNE\cite{Maaten2008VisualizingDU}. The embeddings learned by the MTL version are located in a more reasonable subspace.}

	\begin{figure*}[h!]
	\centering
	\includegraphics[width=1\linewidth]{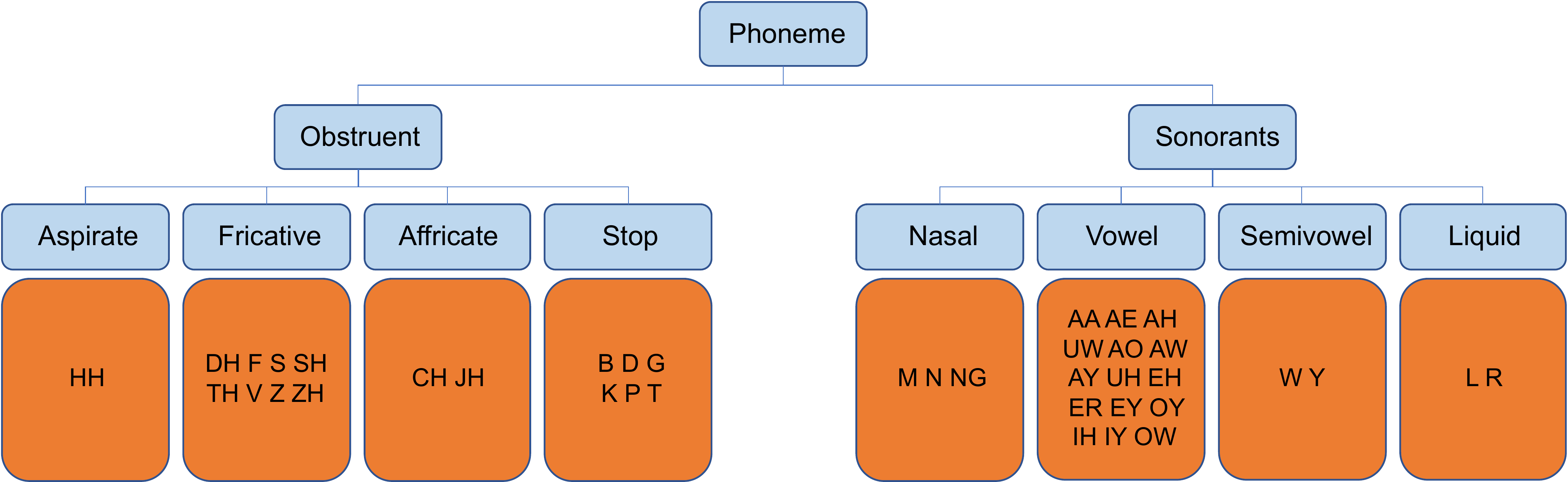}
	\caption{The hierarchy of phonemes in English language.}
	\label{fig:hierarchy}
\end{figure*}

	\begin{figure}[!htbp]
		\centering
		\begin{minipage}[b]{1\linewidth}
			\includegraphics[width=\linewidth]{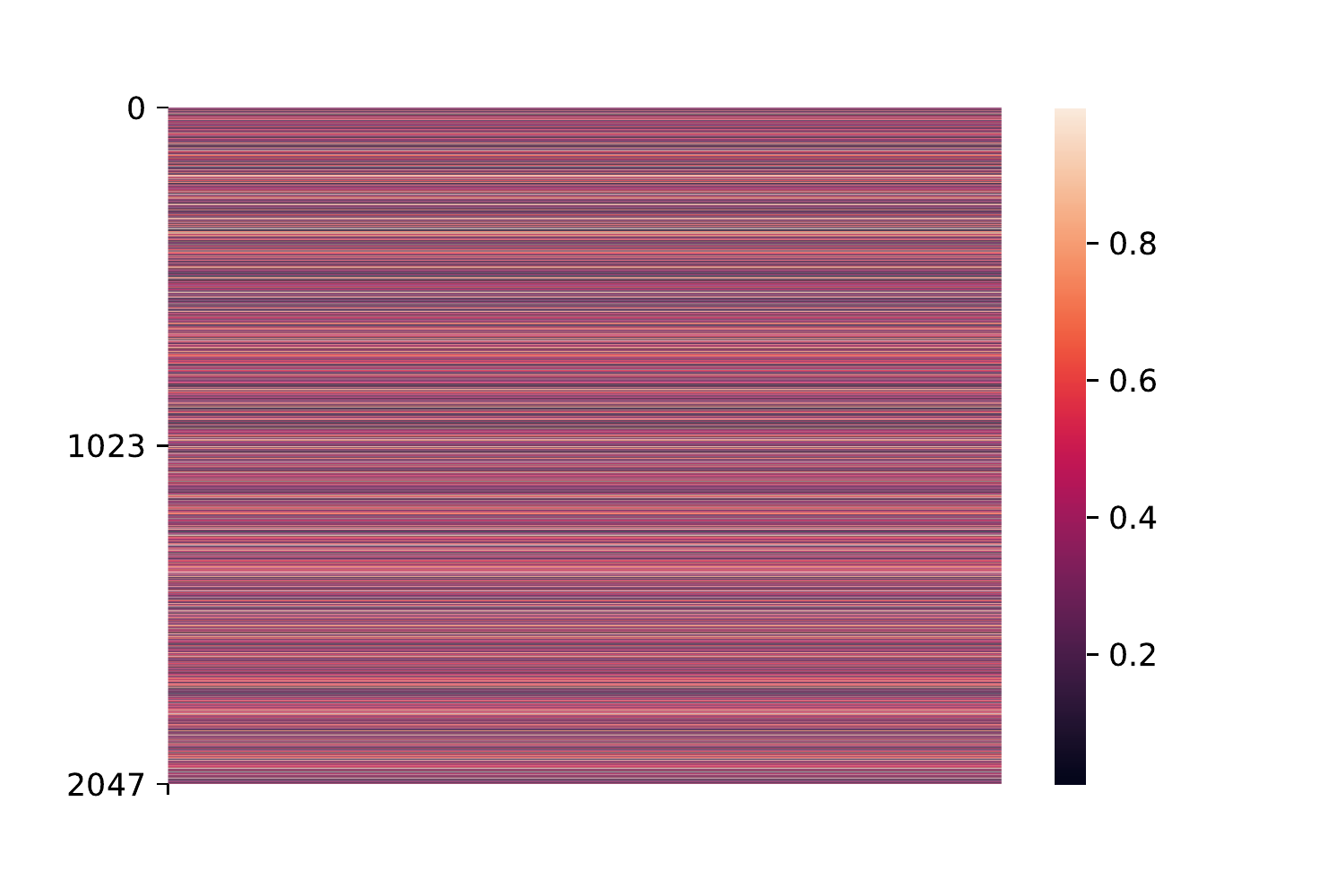}
			\subcaption{ChannelAttention for the model trained with normal transcription. Summed attention weight for the reference embedding dividing summed attention weight for the trainable embedding is $(\sum_{c=d_{emb}}^{2d_{emb}}CA_c)/(\sum_{c=0}^{d_{emb}-1}CA_c)=1.02$.}
			\label{fig:normal_attn}
		\end{minipage}
		\begin{minipage}[b]{1\linewidth}
			\includegraphics[width=\linewidth]{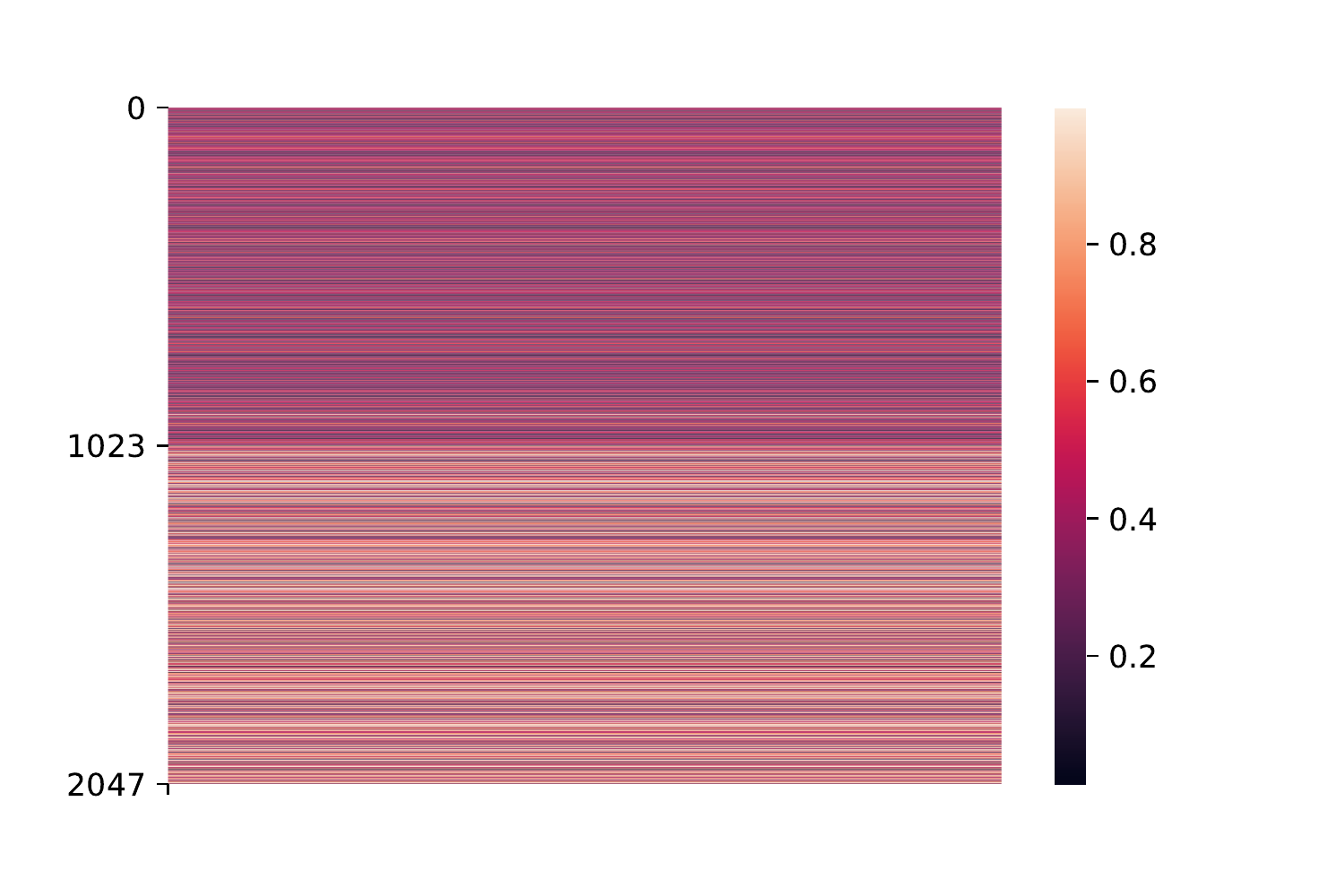}
			\subcaption{ChannelAttention for the model trained with randomly generated transcription. Summed attention weight for the reference embedding dividing summed attention weight for the trainable embedding is $(\sum_{c=d_{emb}}^{2d_{emb}}CA_c)/(\sum_{c=0}^{d_{emb}-1}CA_c)=1.63$.}
			\label{fig:adv_attn}
		\end{minipage}	
		\caption{{Channel attention weight ($CA\in\mathbb{R}^{1\times 2d_{emb}}$) under different training conditions.} When the transcription becomes unreliable, the channel-attention block pays more attention to the reference embedding.}
		\label{fig:attn}
	\end{figure}

	\subsection{Robustness Test}
	As noted in Sec.\ref{sec:hybrid}, if the ASR dataset is constructed by non-native speakers with accents, the transcription of certain phonemes may be labelled as the text that the speakers are asked to read, rather than the pronounced one. In other words, pronunciations with different accents are merged and all labelled as the same text.  If the model is dominated by the auxiliary ASR task, it will be hard for the model to find the uniqueness for different accents. As a result, the performance may be degraded for the AR task. In this subsection, we set experiments to validate the performance of the proposed hybrid model under different ASR situations.
	
	To simulate the transcription confusion introduced by accents, we randomly map a certain phoneme $w$ to the upper-level group it belongs to (denoted as $G(w)$), with the probability $p$ sampled from a uniform distribution $U(0,1)$. The hierarchy of phonemes in English language can be seen in Fig.\ref{fig:hierarchy}, and is commonly used by \cite{Hamooni122014}\cite{Dekel2005}\cite{Lee1989}. Formally, this process can be denoted as follows,
	\begin{equation}
		w'=\begin{cases}
			G(w) &\mbox{if }p<\theta\\
			w &\mbox{otherwise}
		\end{cases},
		p\sim U(0,1).
	\end{equation}
	As $\theta$ increases, there will be more pronunciations that are not labelled as the original phonemes, but are messed into the upper group. {We use the group index for classification instead of the original phoneme index.}
	
	{Meanwhile, to test the effect of the phonetic information for AR, we also test an extreme situation that the text transcriptions are randomly generated. We still use the MTL version (Fig.\ref{fig:mtl}) for experiment, but the trainable acoustic model $AM_t$ is not pretrained on AESRC or Librispeech.  Under this situation, the whole model learns the wrong phonetic information of the accented speech. We also test the proposed hybrid version (Fig.\ref{fig:hybrid}, $AM_t$ is also not pretrained) for comparison.}
	 
	
	
	As we can see from Tab.\ref{tab:rob}, how precise the transcription is indeed has an impact on the AR accuracy. For the simulated dataset, if more phonemes are messed, the performances of both models will be downgraded. {For the MTL version, learning the wrong phonetic information (Id 18) is even worse than the one without ASR pretraining (Id 4 in Tab.\ref{tab:ar_acc}), suggesting that the phonetic information is helpful for the accent recognition.} Meanwhile,  for the proposed hybrid method, the fixed AM can perform the ASR task for the accented speech independently, which is not affected by the messed text, making the model more robust.
    {To see the effect, we plot the channel-attention of the normal situation (Id 13) and the extreme situation (Id 19) in Fig.\ref{fig:attn}. As the random transcription is not helpful for the AR task, the channel-attention mechanism learns to pay more attention to the reference embedding in Fig.\ref{fig:adv_attn} compared to the normal situation in Fig.\ref{fig:normal_attn}.} 

	\section{Conclusion}
	In this paper, we propose a novel hybrid structure along with ASR MTL to solve the AR task. The auxiliary ASR task can force the model to extract {speaker-invariant} and language-related features, which prevents the model from overfitting to the easier speaker recognition task. 
	Furthermore, the hybrid structure is designed to fuse the embeddings of two acoustic models. {The Concatenation-ChannelAttention-based fusion can make the extracted features more robust.}
	The proposed method is 6.57\% better than the official baseline on the validation set and gets a 7.28\% relative improvement on the final test set in the related AESRC competition, showing the merits of our method.
	\label{sec:con}
	\bibliography{refs}
	
\end{document}